\documentclass[prl,twocolumn,showpacs,preprintnumbers,amsmath,amssymb]{revtex4}
\usepackage{graphicx}
\usepackage{bm}
\usepackage{amsmath}
\usepackage{soul}
\usepackage{color}

\begin{document}

\title{Scaling of  geometric phase versus band structure in cluster-Ising models}

\author{Wei Nie\textsuperscript{1}}

\author{Feng Mei\textsuperscript{2}}

\author{Luigi Amico\textsuperscript{1,3,4,5}}

\author{Leong Chuan Kwek\textsuperscript{1,6-8}}

\affiliation{\textsuperscript{1}%
	Centre for Quantum Technologies, National University of Singapore, 3 Science Drive 2, Singapore 117543%
}
\affiliation{\textsuperscript{2}%
	State Key Laboratory of Quantum Optics and Quantum Optics Devices, Institute of Laser Spectroscopy, Shanxi University, Taiyuan, Shanxi 030006, China%
}

\affiliation{\textsuperscript{3} Dipartimento di Fisica e Astronomia, Universit\'a Catania, Via S. Sofia 64, 95123 Catania, Italy}

\affiliation{\textsuperscript{4} CNR-IMM  UOS  Universit\`a  (MATIS), Consiglio  Nazionale  delle  Ricerche $\&$  INFN,  Sezione  di  Catania, Via  Santa  Sofia  64,  95123  Catania,  Italy}

\affiliation{\textsuperscript{5}LANEF {\it 'Chaire d'excellence'}, Universit\`e Grenoble-Alpes \& CNRS, F-38000 Grenoble, France}

\affiliation{\textsuperscript{6}
Institute of Advanced Studies, Nanyang Technological University, 60 Nanyang View, Singapore 639673, Singapore}

\affiliation{\textsuperscript{7}%
National Institute of Education, Nanyang Technological University, 1 Nanyang Walk, Singapore 637616
}

\affiliation{\textsuperscript{8}%
	MajuLab, CNRS-UNS-NUS-NTU International Joint Research Unit, UMI 3654, Singapore%
}

\begin{abstract}
We study the phase diagram of a class of models in which a generalized cluster interaction can be quenched by Ising exchange interaction and external magnetic field. We characterize the various phases through   winding numbers. They may be ordinary phases with local order parameter or exotic ones, known as symmetry protected topologically ordered phases. Quantum phase transitions with dynamical critical exponents $z=1$ or $z=2$ are found. Quantum phase transitions are analyzed through finite-size scaling of the geometric phase accumulated when the spins of the lattice perform an adiabatic precession. In particular, we quantify the scaling behavior of the geometric phase in relation with the topology and low energy properties of the band structure of the system.
\end{abstract}


\maketitle

\textit{Introduction}.$-$
Gapped ground states define quantum phases of matter. Yet, they can be of very different nature. Some of them exhibit approximate orders on a local scale and they can be characterized by their symmetries. Others  possess subtler orders that can only be captured  by highly non-local observables.  One of the major challenging themes in modern condensed matter physics, with applications to quantum technology, is to devise a unified understanding of all the possible quantum phases of matter
\cite{Chen:2011a,Chen:2011b,Kitaev:2009,Sachdev:2011,Savary:2017}.  Indeed, the scientific community has been applying integrated methods by combining quantum information, foundational notions  of quantum mechanics and many-body physics to study the problem \cite{Amico:2008,Eisert:2010,Zeng:2015,Kuwahara:2017}. Recent outcomes  in topological matter e.g., topological insulators, Weyl semi-metals \cite{Xu:2015,Kondo:2015,Bansil:2016} and superconductors have demonstrated how  the topology of the energy bands of the system can be useful in a novel way for analyzing the quantum phases of matter \cite{Bansil:2016,Franz:BOOK}. Here, we study  a specific many-body system that can  display exotic orders by exploiting Berry phase and winding numbers \cite{Berry:1984,Anderson:1958,Thouless:1982}.

We focus on a one dimensional spin system whose ground state can be tuned to be an ordered state with local order parameter or  to be a state with an exotic order of topological nature (so-called symmetry protected topological order \cite{Chen:2013,Pollmann:2009}).
The different regimes that may be established in the system are separated by  quantum phase transition driven by certain control parameters \cite{Sachdev:2011,Sachdev:BOOK}.
The criticality between ground states with exotic order has been  receiving an outgrowing interest \cite{Lahtinen:2015,Ohta:2016}. Interestingly, it was found that the Berry phase can be invoked to study quantum phase transition \cite{Carollo:2005,Zhu:2006,Hamma:2006,Peng:2010}.  Finite-size scaling analysis of geometric phase to reveal order-disorder  quantum phase transition was studied in the $XY$ spin chain \cite{Zhu:2006}.

In this paper, we consider a set of   localized spins in one dimensional lattice enjoying a  specific higher order (multispin) interaction; at the same time, such interaction competes with Ising exhange; finally the chain is placed in a transverse magnetic field. Indeed, the systems under scrutiny  are  generalizations of the cluster-Ising model that was formulated in the cross-fertilization area between many-body physics, quantum correlations and ultracold atoms \cite{Skrovseth:2009,Son:2011,Pachos:2004}. The cluster-Ising model  displays a second order quantum phase transition between a phase with local order parameter and a symmetry protected topological quantum phase \cite{Smacchia:2011,Montes:2012}. Here, we  study the phase diagrams of the generalized cluster-Ising models by looking at the winding numbers of the ground states  (for a specific subclass of models studied here, winding numbers were recently studied in \cite{Zhang:2015}). We explore the criticality of the system by studying the finite-size scaling of the geometric phase. In particular, we find that critical points with nonlinear low-energy dispersions are characterized by an anomalous logarithmic scaling of the geometric phase.

\begin{figure}[!htbp]
\includegraphics[width=8cm]{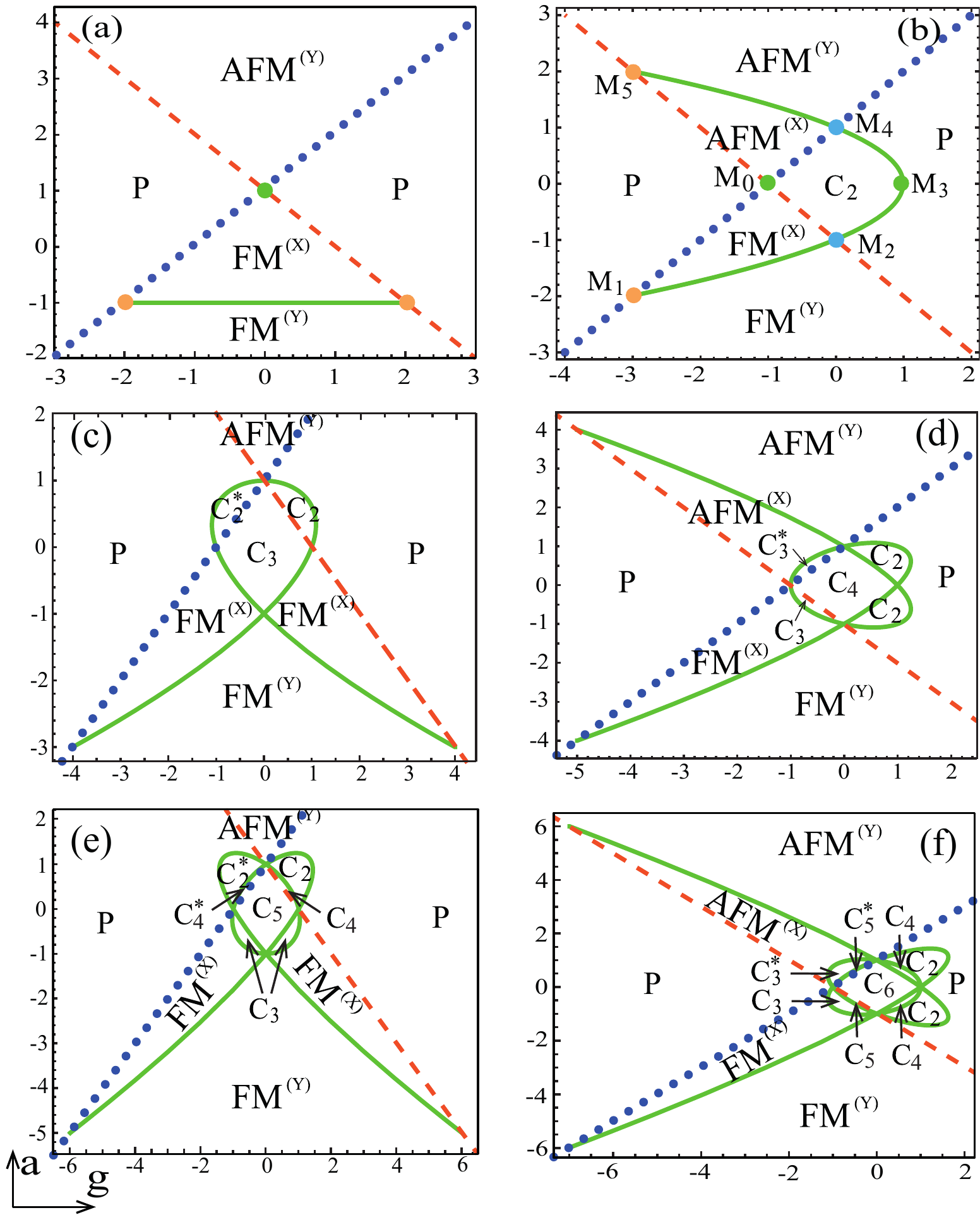}
\caption{(Color online). Phase diagrams of cluster-Ising models.  (a)-(f) are phase diagrams of the system for $l$ from $1$ to $6$. We choose $\lambda=1$. The horizontal and vertical axes are represented by $g$ and $a$, respectively. The abbreviations mean different phases: paramagnetic ($\mathrm{P}$), ferromagnetic ($\mathrm{F}$), antiferromagnetic ($\mathrm{AF}$), cluster ($\mathrm{C}$); the superscripts $\mathrm{X},\mathrm{Y}$ specify the directions of the orders. $C_{l}$ represent the cluster phase ($l=2$) or generalized cluster phases ($l>2$). $C_{l}^{\ast}$ mean  dual (generalized) cluster phases (see the text for the definition).
}\label{phasediagram1}
\end{figure}
\textit{Model.}$-$We consider a class of models describing  interactions  between $l+1$ spins competing with exchange interaction,  in external field. The Hamiltonian reads \cite{Lahtinen:2015,Ohta:2016,Skrovseth:2009,Son:2011,Pachos:2004,Smacchia:2011,Montes:2012}
\begin{eqnarray}
H^{(l)} =  \sum_{j=-M}^{M} -\lambda \sigma_{j}^x {\cal{Z}}_{j,l} \sigma_{j+l}^x  + a \sigma_j^{y} \sigma_{j+1}^y + g  \sigma_j^z,
\label{H1}
\end{eqnarray}
with ${\cal{Z}}_{j,l}=\sigma_{j+1}^z...\sigma_{j+l-1}^z$ and $M=(L-1)/2$ for odd $L$. The operators $\sigma_n^{\alpha} (\alpha=x,y,z)$ are the Pauli matrices defining the spin state in the $n$-th site  of the one dimensional lattice. Eq. (\ref{H1}) can be mapped to a system of decoupled $l+1$ free fermions: $H^{(l)} = \sum_{k} \Psi_k^{\dagger} H^{(l)}_{k} \Psi_k$ where   $\Psi_k^{\dagger} = (c_k^{\dagger}, c_{-k})$  and $H^{(l)}_{k} = \mathbf{d}^{(l)}(k) \cdot \bm{\sigma}$  lies in a pseudo-spin Hilbert space, with $\mathbf{d}^{(l)}(k) = h^{(l)}_y \hat{\mathbf{e}}_y + h^{(l)}_z \hat{\mathbf{e}}_z$, $h^{(l)}_y =\lambda \sin k l +a \sin k$, $h^{(l)}_z =-\lambda \cos k l + a \cos k-g$, being  $\hat{\mathbf{e}}_y$, $\hat{\mathbf{e}}_z$ the unit vectors in the directions $y,z$.

For  $l=1$,  $Z_{j,l}=1$,  and therefore the Hamiltonian defines the transverse Ising model with the well-known  antiferromagnet-paramagnet quantum phase transition in the Ising universality class. For $l=2$,  Eq. (\ref{H1}) defines  the cluster-Ising model in an external magnetic field.  Assuming periodic boundary conditions: $\sigma_{L+1}^{\alpha} = \sigma_1^{\alpha}$, the ground state of Eq. (\ref{H1}) for $a=g=0$ is a unique state known as cluster state \cite{Raussendorf:2003}. Such state enjoys a non-trivial global symmetry of the $\mathbb{Z}_2 \times \mathbb{Z}_2$ type. For open boundary conditions, the cluster state  is fourfold degenerate. Such a degeneracy can be lifted only by resorting to operators in  the Hamiltonians's symmetry algebra.  In  such a specific sense, the cluster ground state provides  an example of quantum phase of matter with the symmetry protected topological order \cite{Chen:2013}.  Remarkably,  such a kind of order is  preserved by the  Ising interaction and  the external  field in Eq. (\ref{H1})  until  quantum phase transitions occur into the system. The cluster-Ising models enjoy non-trivial  duality properties \cite{Savit:1980,Son:2012,Son:2011,Smacchia:2011,Ohta:2016,Skrovseth:2009,Savary:2017}.  In particular, our Hamiltonians Eq. (\ref{H1}) can be mapped to the class of models considered in Ref. \cite{Ohta:2016} by $\sigma_j^z = \tau_j^y \tau_{j+1}^y$,
$\sigma_j^y \sigma_{j+1}^y = -\tau_j^y \tau_{j+1}^z\tau_{j+2}^y$,
$\sigma_{j-1}^x \sigma_j^z \sigma_{j+1}^z \sigma_{j+2}^x = -\tau_j^x \tau_{j+1}^z \tau_{j+2}^x$. Since the construction of the phase diagram of the systems relies on  energy properties,  the phase diagrams are unaltered  by duality.

\begin{table}[b]
  \caption{Phase and winding number for specific part in the cluster-Ising models.}
  \label{table}
\begin{ruledtabular}
  \begin{tabular}{ccc}
    Interaction  & Phase & Winding number    \\
    \hline
    $\pm \sum_j \sigma_j^z$  & $\mathrm{P}$ & $0$   \\
    $\pm \sum_j \sigma_j^y \sigma_{j+1}^y$  & $\mathrm{AFM}^{(\mathrm{Y})}, \mathrm{FM}^{(\mathrm{Y})}$ & $+1$   \\
    $\pm \sum_j \sigma_j^x \sigma_{j+1}^x$ & $\mathrm{AFM}^{(\mathrm{X})}, \mathrm{FM}^{(\mathrm{X})}$ & $-1$   \\
    $\pm \sum_j \sigma_{j}^x {\cal{Z}}_{j,l} \sigma_{j+l}^x$   & $C_{l}^{\ast}$, $C_{l}$ & $-l$   \\
  \end{tabular}
\end{ruledtabular}
\end{table}

\textit{Phase diagram.}$-$
Different ground states can be characterized by order parameters. In the cluster phases, however such order parameters need to be highly non-local (the string order parameters) \cite{Smacchia:2011,Ohta:2016}. Winding numbers provide an alternative description  of the different phases bypassing the notion of order parameter \cite{Anderson:1958}.  Indeed, winding numbers have integer values and they cannot change without closing the spectral gap. Remarkably, these numbers correspond to the number of zero modes appearing at the edge of the system (when open boundary conditions  are imposed).

Winding number counts the times that a closed curve encircles the origin in the pseudo-spin Hilbert space $(h^{(l)}_y(k), h^{(l)}_z(k))$:
\begin{equation}
W=\frac{1}{2\pi}\int_{\mathrm{B}.\mathrm{Z}.} d\theta^{(l)}_k,
\end{equation}
with $\theta^{(l)}_k=\mathbf{d}^{(l)}(k)/|\mathbf{d}^{(l)}(k)|$. 
The winding numbers in the different ground states of the system are summarized in the Table \ref{table}. $\mathrm{FM}^{\alpha}$ ($\mathrm{AFM}^{\alpha}$) denotes ferromagnetic (antiferromagnetic) order along the spin direction $\alpha$.

The cluster order can be quenched by local field $g$ and nonlocal exchange interaction $a$ \cite{Bravyi:2010, Brown:2011}.
\begin{figure}[t]
\includegraphics[width=8cm]{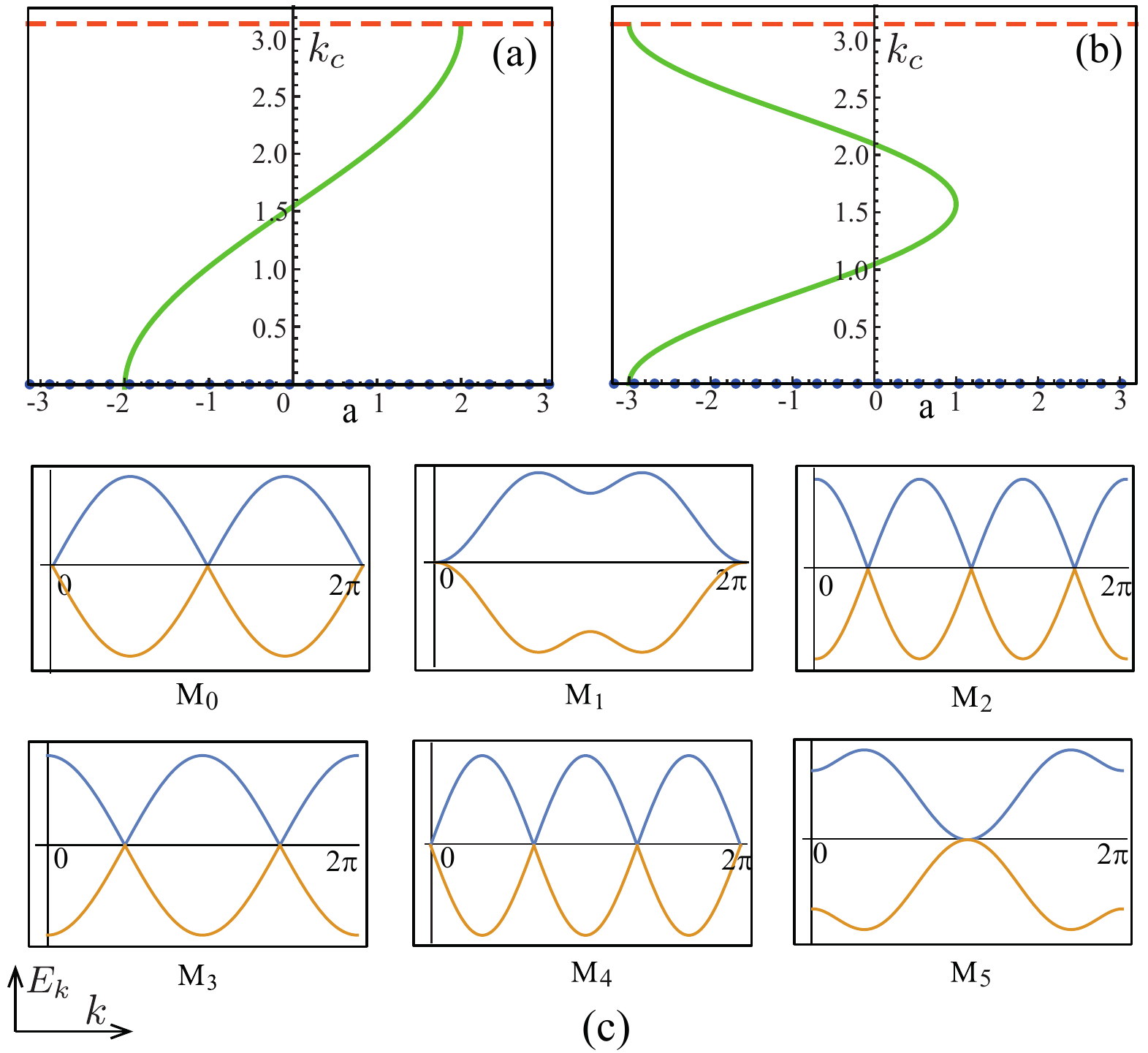}
\caption{(Color online). (a) and (b) are the critical momentum corresponding to $l=2$ and $3$, respectively. (c) Energy band structures of critical points $M_0$-$M_5$ for $l=2$ in Fig. \ref{phasediagram1}(b). $M_1$ and $M_5$  are non-Lorentz-invariant critical points, with nonlinear low-energy modes.  In $M_0$ and $M_3$ the band structure share the same topology. Similarly for multicritical points $M_2$ and $M_4$.
}\label{energyband}
\end{figure}
In Figs. \ref{phasediagram1}(a)-\ref{phasediagram1}(f) we show the phase diagrams with $l$ from $1$ to $6$. (a) and (b) show the detailed phase diagrams for $l=1$ and $2$ (see Refs. \cite{Zhang:2015,Ohta:2016}). The cases $l>2$ were recently studied by  Lahtinen and Ardonne \cite{Lahtinen:2015}.

We observe that  the generalized cluster states with winding number $-l$ are ``broken" into phases characterized by winding numbers $(-(l-1),\ldots,-1)$.
The structure of  the phase diagrams  is related to the symmetry of  $(h_y(k), h_z(k))$. Such symmetry implies the parity of the number of $l$. For even $l$ (Figs. \ref{phasediagram1}(b), \ref{phasediagram1}(d) and \ref{phasediagram1}(f)), the Zeeman field is the control parameter. When $g>0$, phases with even integer winding numbers are generated. The Ising interaction $a$ tunes the ferromagnetic or antiferromagnetic phase. The roles of $g$ and $a$ are exchanged for odd $l$ (Figs. \ref{phasediagram1}(a), \ref{phasediagram1}(c) and \ref{phasediagram1}(e)). The  $C_2$ and $C_2^{\ast}$ (the so-called dual  cluster phase) display a string order of the cluster state type with two Majorana modes at the edges of the system; such two phases are characterized by string order parameters with different spin polarizations at the edges \cite{Ohta:2016}. Similarly, $C_3$ phases are cluster phases with three Majorana modes at the edges of the system.
 $C_3$ and $C_3^{\ast}$ phases in Fig. \ref{phasediagram1}(d) are distinguished from each other by the negative and positive Ising interaction $a$.
The $C_m$ and $C_m^*$ phases with $m>2$  in Figs. \ref{phasediagram1}(c)-\ref{phasediagram1}(f)  are defined  with a similar logic. The different phases $C_m$  with fixed $l$ in the different panels of Fig. \ref{phasediagram1} can be connected adiabatically:  a fixed phase $C_m$ of a given Hamiltonian $H^{(l)}$ evolves into $C_m$ of $H^{(l+1)}$ under $H^{(l,l+1)}=\left ( t-1\right ) H^{(l)}+t H^{(l+1)}$, $t\in [0,1]$.

Phase boundaries  are obtained as the combination of the Hamiltonian parameters for which a specific low-energy mode emerges (for a specific value of critical momentum $k_c$) in the band structure. The critical momenta  of the phase boundaries  Figs. \ref{phasediagram1}(b) and \ref{phasediagram1}(c) are shown in Figs. \ref{energyband}(a) and \ref{energyband}(b), respectively. The green-solid lines in Figs. \ref{phasediagram1} are in the $XX$ universality class. Along there, the   critical momentum depends on the parameters $a$ and $g$. For the blue-dotted (red-dashed) straight lines indicating Ising phase transitions in Fig. \ref{phasediagram1}, there is one Dirac point at $k_c=0$ ($\pi$). The $XX$ and Ising transitions have a topological difference. The two phases separated by the $XX$ line have  winding number difference equals to  $2$. However, for ground states separated by the Ising type transition, the winding number difference is $1$. Elaborating on the findings for $l=2$ \cite{Smacchia:2011}, Lahtinen and Ardonne demonstrated that the multicritical points  of the system may be  indeed characterized by  the $so(N)_1$ conformal field theory. For $M_0$ with two Ising criticalities, the $XX$ gets to the  $so(2)_1$ universality class. If the cluster type of order is involved,  more branches ($\geq 3$) with linear dispersions show up at the criticality (Fig. \ref{energyband}(c)). The multicritical points are combined in the $so(l)_1$ and $so(l+1)_1$ universality classes \cite{Lahtinen:2015}. Specifically, in $M_3$ there are two degenerate points in Brillouin zone.  As for  $M_4$, there are three degenerate points in the band structure. Therefore, $M_4$ enjoys a  $so(3)_1$ criticality rather than  the $XX$ one. It is interesting to note that there is quadratic band touching at $M_1$ and $M_5$. Indeed, a quadratic band touching may lead to interesting non-Fermi liquid interaction effects \cite{Moon:2013,Herbut:2014}.

\begin{figure*}[!htbp]
\begin{center}
\includegraphics[width=16cm]{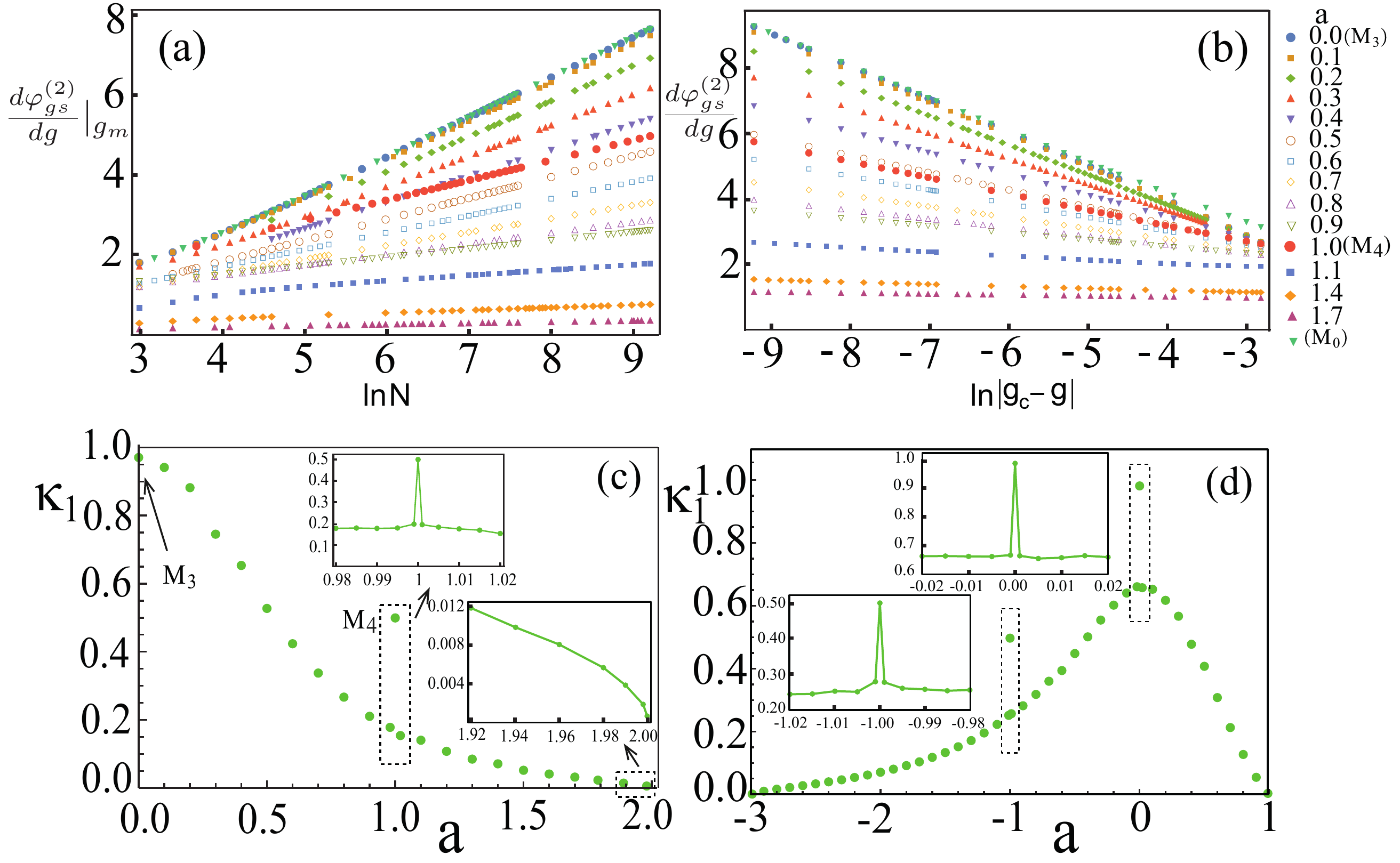}
\end{center}
\caption{(Color online). (a),(b): The scaling behavior  close to the $XX$ and $so(3)_1$ ($M_4$) criticalities in Fig. \ref{phasediagram1}(b) with various value of $a$ {( indicated at the right side of (b)}). (c) displays the scaling parameter $\kappa_1$ shown in (a) with $ a \in [0,2)$. (d) denotes $\kappa_1$ of $XX$ (the part with $k_c$ changing from $0$ to $\pi/2$), $so(3)_1$ and $so(4)_1$ criticalities for $l=3$ (Fig. \ref{phasediagram1}(c)) with $ a \in (-3, 1]$. A sudden jump of $\kappa_1$ marks the transition between two nearly gapless phases with different topologies.}
\label{scaling1}
\end{figure*}
\textit{Scaling of geometric phase.}$-$
Berry phase arises when the spin variables localized in the lattice points along the chain are rotated adiabatically \cite{Carollo:2005}. The rotating system can be described by $H^{(l)}_{\mathcal{R}} = \mathcal{R}^{\dagger} H^{(l)} \mathcal{R}\quad$, with $\mathcal{R}^{\dagger}= e^{i \sum_{j=-M}^{M} \phi \sigma_j^z/2}$. For our model Eq. (\ref{H1}), the ground state of $H^{(l)}_\mathcal{R}$ is the vacuum of free fermionic modes: $|gs\rangle ^{(l)}=\prod_{k=1}^M |gs\rangle^{(l)}_k$ with $|gs\rangle^{(l)}_k=(\cos\frac{\theta^{(l)}_k}{2}|0\rangle_k |0\rangle_{-k} - i e^{-i 2\phi} \sin\frac{\theta^{(l)}_k}{2}|1\rangle_k|1\rangle_{-k} )$ where $|0\rangle_k$,  $|1\rangle_k$ are the vacuum and single excitation of the $k$-th mode, $c_k$, respectively.
Adiabatically varying the angle $\phi$ from $0$ to  $\pi$, the geometric phase of  $|gs\rangle$  results \cite{Carollo:2005,Zhu:2006}
\begin{eqnarray}
\varphi_{gs}^{(l)}&=&{\frac{i}{M}}\int_0^{\pi}{}^{(l)}\langle gs|\partial_{\phi} |gs\rangle^{(l)} d \phi \nonumber \\
&=&{\frac{\pi}{M}}\sum_{k}(1-\cos \theta^{(l)}_k).
\end{eqnarray}

In Fig. \ref{phasediagram1}(a) the horizontal line (green-solid) at $a=-1$, $-2 < g < 2$   defines a  $XX$ critical state with quasi-long-range order. In such a state, the Berry phase is identically vanishing.
If a non-trivial cluster state order is involved the universality class of the transition changes: multicritical points $M_2$ and $M_4$ appear as shown in Fig. \ref{phasediagram1}(b). The Berry phase near  the criticality (green-solid) is non-vanishing. We explore the criticality  via $d \varphi/ d g$.
\\
{\it (I)} Scaling close to quantum phase transitions with the {critical exponent} $z=1$.
The scaling ansatz for (derivative of ) the geometric phase is \cite{Zhu:2006}
\begin{equation}
\frac{d \varphi_{gs}^{(l)}}{d g}|_{g_m} \simeq \kappa_1 \ln N + \mathrm{const},
\label{scalingN}
\end{equation}
\begin{equation}
\frac{d \varphi_{gs}^{(l)}}{d g} \simeq \kappa_2 \ln |g-g_c| + \mathrm{const},
\label{scalingg}
\end{equation}
where $g_c$ is the critical value of $g$ for infinite long spin chain, and $g_m$ marks the anomaly for  the finite-size system. According to the scaling ansatz, in the case of logarithmic singularities, the ratio $|\kappa_2 / \kappa_1|$ is the exponent $\nu$ that governs the divergence of correlation length. We note that the scaling behavior is related to the band structure at low energy.

\begin{figure}[t]
\includegraphics[width=8cm]{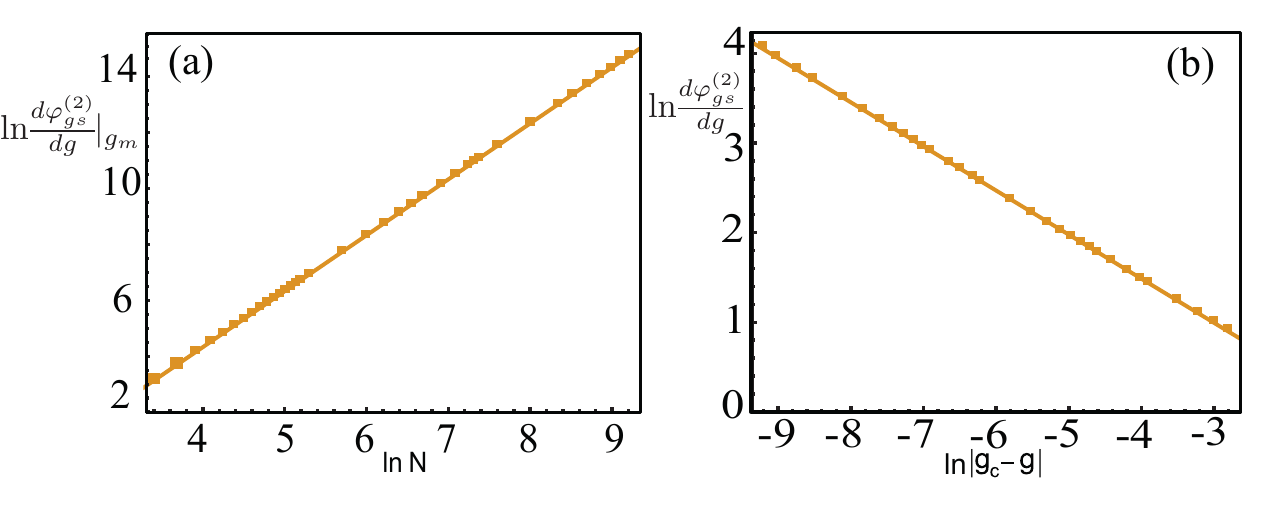}
\caption{(Color online). (a) and (b) display scaling behavior at the point $M_5$.}
\label{scaling2}
\end{figure}

As for  topological quantum phase transitions, we first consider $l=2$. The critical properties are found symmetric about $a=0$ as shown in Fig. \ref{phasediagram1}(b).  We discuss the phase boundaries with $0\leq a \leq 2$. For  $a=0$, phase transitions occur at $|g|=1$ which separates a paramagnetic and a cluster phase. As expected by looking at the dispersion curves,  $M_0$ and $M_3$ share the same criticality. Similarly, the quantum multicritical points $M_2, M_4$ involving the cluster phase enjoy the same scaling behavior. In Figs. \ref{scaling1}(a) and \ref{scaling1}(b) we present the scaling behaviors characterized by Eqs. (\ref{scalingN}) and (\ref{scalingg}). For critical regime with critical exponent $z=1$  and linear low-energy dispersions, the ratio $|\kappa_2/\kappa_1| \sim 1$. The scaling parameter $\kappa_1$ for $l=2$ is represented in (c). We find that for multicritical points with multiple degeneracies in energy bands, the scaling coefficients are discontinuously connected to the neighboring critical points which share the same topologies of band structures.  The discontinuity (sudden change) of the scaling parameter $\kappa_1$ renders the topological change of band structure. The smooth variation of $\kappa_1$ in the $XX$ criticality arises from the fact that slopes of linear dispersions change depending on Ising exchange interaction and transverse magnetic field. In Fig. \ref{scaling1}(d), we show $\kappa_1$ for $l=3$ and also observe the sudden change at multicritical points. Similar behaviors also exist for $\kappa_2$.
\\
{\it (II) } Scaling close to quantum phase transitions with $z=2$. Quantum phase transitions implied in Eq. (\ref{H1}) may be  characterised by a low-energy dispersions $\sim k^2$ (e.g., $M_1$ and $M_5$ for $l=2$).  Consistently with the scaling theory,  the dynamical critical index for such phase transition is $z=2$. The scaling behavior of this band topology at critical regime are shown in Fig. \ref{scaling2}. For $M_1$ and $M_5$,  we found that the scaling ansatz  Eqs. (\ref{scalingN}) and (\ref{scalingg}) should be modified to
\begin{equation}
\ln \frac{ d \varphi_{gs}^{(l)}}{d g}|_{g_m} \simeq \tilde{\kappa}_1 \ln N + \mathrm{const},
\end{equation}
and
\begin{equation}
\ln \frac{ d \varphi_{gs}^{(l)}}{d g} \simeq \tilde{\kappa}_2 \ln |g-g_c| + \mathrm{const}.
\end{equation}
$\tilde{\kappa}_1$, and $\tilde{\kappa}_2$ in Figs. \ref{scaling2}(a) and \ref{scaling2}(b) are found $1.999$ and $-0.492$, respectively.  Close to critical points with quadratic dispersions for  $l > 2$,  we find a similar log scaling behavior.

\textit{Conclusions.}$-$ We have  studied the criticality of generalized cluster-Ising models through the scaling properties of the geometric phase. The criticality  with parameter-dependent critical momentum is generically found  of the $XX$ type. At the multicritical points with linear gapless modes the quantum phase transitions are in the  $so(N)_1$ universality classes. We have found that the critical points with linear and quadratic low-energy dispersions obey different scaling ansatz.  Specifically, the critical point with critical exponent $z=2$ shows anomalous logarithmic scaling behavior which is markedly  different from that one  with  $z=1$, with linear dispersions. We also employed the scaling parameters to study the band topology in critical regime. We have also found that the scaling parameters change smoothly along the phase boundary with $z=1$. In contrast, the scaling parameters are found very sensitive to topological change (Figs. \ref{scaling1}(c) and \ref{scaling1}(d)). In this paper, we observed  that there is a close connection between topological phase transition, quantum criticality, energy band structure and geometric phase.

Our approach may be generalized to other spin chain with multispin interactions, like  the  Wen-plaquette model, simulated in nuclear magnetic resonance systems, \cite{Peng:2014}, or Baxter-Wu models \cite{Baxter:1973,Penson:1982}. Recently,  multispin interactions were demonstrated to arise in Floquet driven lattices \cite{Benito:2014,Potirniche:2016}.
We finally observe that the multicritical points in the generalized cluster-Ising models can be used to investigate nonequilibrium dynamics in many-body physics \cite{Tomka:2012,Heyl:2013}. It is interesting to study the non-adiabatic driving scheme across the multicritical points in cluster-Ising spin chain and the interplay between geometric phase and dynamics in the near future.

\begin{acknowledgments}
W.N. would like to thank V.M. Bastidas and Ching Hua Lee for useful discussions. F.M. is supported by the National Natural Science Foundation of China (grant no. 11604392). The Grenoble LANEF framework (ANR-10-LABX-51-01) is acknowledged for its support with mutualized infrastructure. L.C.K. etc acknowledge support from the National Research Foundation \& Ministry of Education, Singapore.
\end{acknowledgments}

\bibliographystyle{apsrev}

\end{document}